\documentclass[twoside]{article}
\usepackage{fleqn,espcrc2}

\usepackage{epsfig}
\usepackage{graphicx}
\usepackage[figuresright]{rotating}

\newcommand{\be}{\begin{equation}}
\newcommand{\ee}{\end{equation}}
\newcommand{\ba}{\begin{eqnarray}}
\newcommand{\ea}{\end{eqnarray}}

\newcommand{\AmS}{{\protect\the\textfont2
  A\kern-.1667em\lower.5ex\hbox{M}\kern-.125emS}}
\def\spose#1{\hbox to 0pt{#1\hss}}
\def\ltapprox{\mathrel{\spose{\lower 3pt\hbox{$\mathchar"218$}}
 \raise 2.0pt\hbox{$\mathchar"13C$}}}
\def\ad#1{$\,^{\rm #1}$}

\def\p{^\prime}

\title{Comparison of finite-size-scaling functions for $3d$ $O(N)$ spin models\\
        to QCD \thanks{Talk by T. Schulze, we thank for supporting grants
       DFG, No. FOR 339/1-2 and FAPESP, Brazil, No. 00/05047-5.} }

\author{T. Schulze\address{Fakult\"at f\"ur Physik, Universit\"at Bielefeld,
    D-33615 Bielefeld, Germany} with J. Engels\ad a,
    S. Holtmann\ad a  and T. Mendes\address{IFSC-USP, Caixa postal 369,
    13560-970 S\~ao Carlos SP, Brazil}}

\begin{document}

\begin{abstract}
We calculate numerically universal finite-size-scaling functions of the
magnetization for the three-dimen\-sio\-nal $O(4)$ and $O(2)$ spin models. The
approach of these functions to the infinite-volume scaling functions is studied
in detail on the critical and pseudocritical lines. For this purpose we
determine the pseudocritical line in two different ways. We find that 
the asymptotic form of the finite-size-scaling functions is already 
reached at small values of the scaling variable. A comparison with QCD 
lattice data for two flavours of staggered fermions shows a similar 
finite-size behaviour which is compatible with that of the spin models.
\vspace{1pc}
\end{abstract}

\maketitle

\section{INTRODUCTION}

Our motivation for studying $3d$ $O(N)$ spin models is their relation 
to quantum chromodynamics (QCD). The QCD chiral phase transition 
for two light-quark flavors is supposed to be of second order in 
the continuum limit and to be in the same universality 
class as the $3d$ $O(4)$ model \cite{PW}. In the staggered formulation
of QCD on the lattice a part of chiral symmetry is remaining and that is 
$O(2)$. The comparison of QCD lattice data to $O(4)$ scaling functions for $V
\rightarrow \infty$ and exponents yields good agreement for Wilson
fermions \cite{CPPACS}, but not for staggered fermions \cite{MILC}. This may be
caused by using too small lattices in QCD, so we will compare in this paper
staggered QCD data to the finite-size-scaling functions of the 3d $O(2)$ and $O(4)$
spin models.\\
\indent The $O(N)$-invariant nonlinear $\sigma$-model on a $d-$
dimensional hypercubic
lattice is defined by
\be
\beta\,{\cal H}\;=\;-J \,\sum_{<i,j>} {\bf S}_i\cdot {\bf S}_j
         \;-\; {\bf H}\cdot\,\sum_{i} {\bf S}_i \;.
\ee
${\bf S}_i$ is an N-component unit vector at site $i$
with a longitudinal (parallel to the magnetic 
field ${\bf H}$) and a transverse component 
\be
{\bf S}_i\; =\; S_i^{\parallel} {\bf \hat H} + {\bf S}_i^{\perp} ~.
\ee
The order parameter of the system, the magnetization $M$, is given by
\be
M \;=\, \langle\frac{1}{V}\sum_{i} S_i^{\parallel}\,\rangle\; =\; \langle\,
S^{\parallel}\,\rangle~,
\ee
and the longitudinal susceptibility by
\be
\chi_L\;=\; {\partial M \over \partial H}
 \;=\; V(\,\langle\, { S^{\parallel}}^2\rangle\, -M^2)~. \label{chil}
\ee


\section{FINITE-SIZE-SCALING FUNCTIONS}

\medskip
The general form of the finite-size-scaling function for the 
magnetization is given by
\be
M = L^{-\beta/\nu} \Phi(tL^{1/\nu},hL^{1/\nu_c},L^{-\omega})~,
\ee
where $t$ and $h$ are the normalized reduced temperature $t=(T-T_c)/T_0$ and
magnetic field $h=H/H_0$. On lines of fixed $z=th^{-1/\beta\delta}$ and after
expanding $\Phi$ in $L^{-\omega}$ we have
\be
M = L^{-\beta/\nu} Q_z(hL^{1/\nu_c})+\dots ~,
\ee
where $Q_z$ is a {\em universal} function. Examples of lines of fixed $z$
are the critical line ($z=0$) and the pseudocritical line ($z=z_p$), the line
of maximum positions of the susceptibility $\chi_L$ in the $(t,h)$-plane
for $V\rightarrow \infty$.

For large lattices $Q_z$ will converge to the asymptotic form $Q_{z,\infty }$
\be
Q_z\rightarrow Q_{z,\infty }\;=\;f_G(z)(hL^{1/\nu_c})^{1/\delta}~.
\ee
$Q_{z,\infty }$ can be calculated from the scaling function $f_G$ for
$V\rightarrow \infty$, which we have calculated in Ref. \cite{O4} for O(4) and
in Ref. \cite{O2} for O(2).
\begin{figure}[t!]
  \epsfig{bbllx=97,bblly=290,bburx=421,bbury=587,
        file=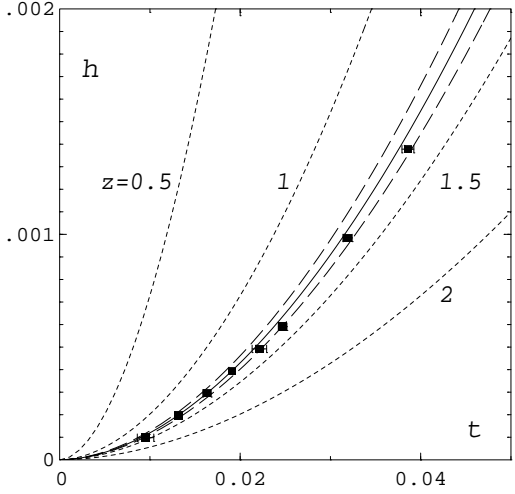, width=60mm}
\caption{Lines of fixed $z=0.5, 1, 1.5, 2$ (dashes), the pseudocritcal line
        (solid) at $z_p=1.33\pm 0.05$ and from measured peak positions
        (squares).}
\label{fig:z_plineO4}
\end{figure}


\section{NUMERICAL RESULTS}
\label{section:numres}

\medskip
Our simulations were done on three-di\-men\-sio\-nal lattices with periodic 
boundary conditions and linear extensions $L=8-120$ using the cluster 
algorithm. In order to eliminate corrections to scaling we used increasingly
larger values of $L$. For the critical coupling and the exponents we took the
same values as in Refs. \cite{O4} and \cite{O2}.
  
First we had to determine the value $z_p$ of the pseudocritical line. The
scaling function of $\chi_L$ for $V\rightarrow \infty$ can be determined from
the scaling function of the magnetization $f_G$ 
\be
\chi_L={\partial M\over \partial H}={h^{1/\delta-1} \over H_0 \delta}
\left ( f_G(z) - {z\over \beta}f_G\p(z) \right)~.
\label{max}
\ee

Evidently, the maximum of $\chi_L$ at fixed $h$ and varying $t$ is at the
maximum point $z_p$ of the function in the brackets of Eq.\ (\ref{max}).
We obtain
$z_p=1.56(10)$ for $O(2)$ and $z_p=1.33(5)$ for $O(4)$.

As a check we have also determined the peak positions on lattices with $L=24-96$
which extrapolate for $L\rightarrow \infty$ to $z_p=1.65(10)$ for $O(2)$ and
$z_p=1.35(10)$ for $O(4)$. In Fig.\ \ref{fig:z_plineO4} we compare both methods
for the $O(4)$ model. We see that both methods agree very well for small
$h$, but for larger $h$ the extrapolated peak positions yield slightly higher
pseudocritical temperatures.

\begin{figure}[t!]
   \epsfig{bbllx=40,bblly=290,bburx=370,bbury=560,
       file=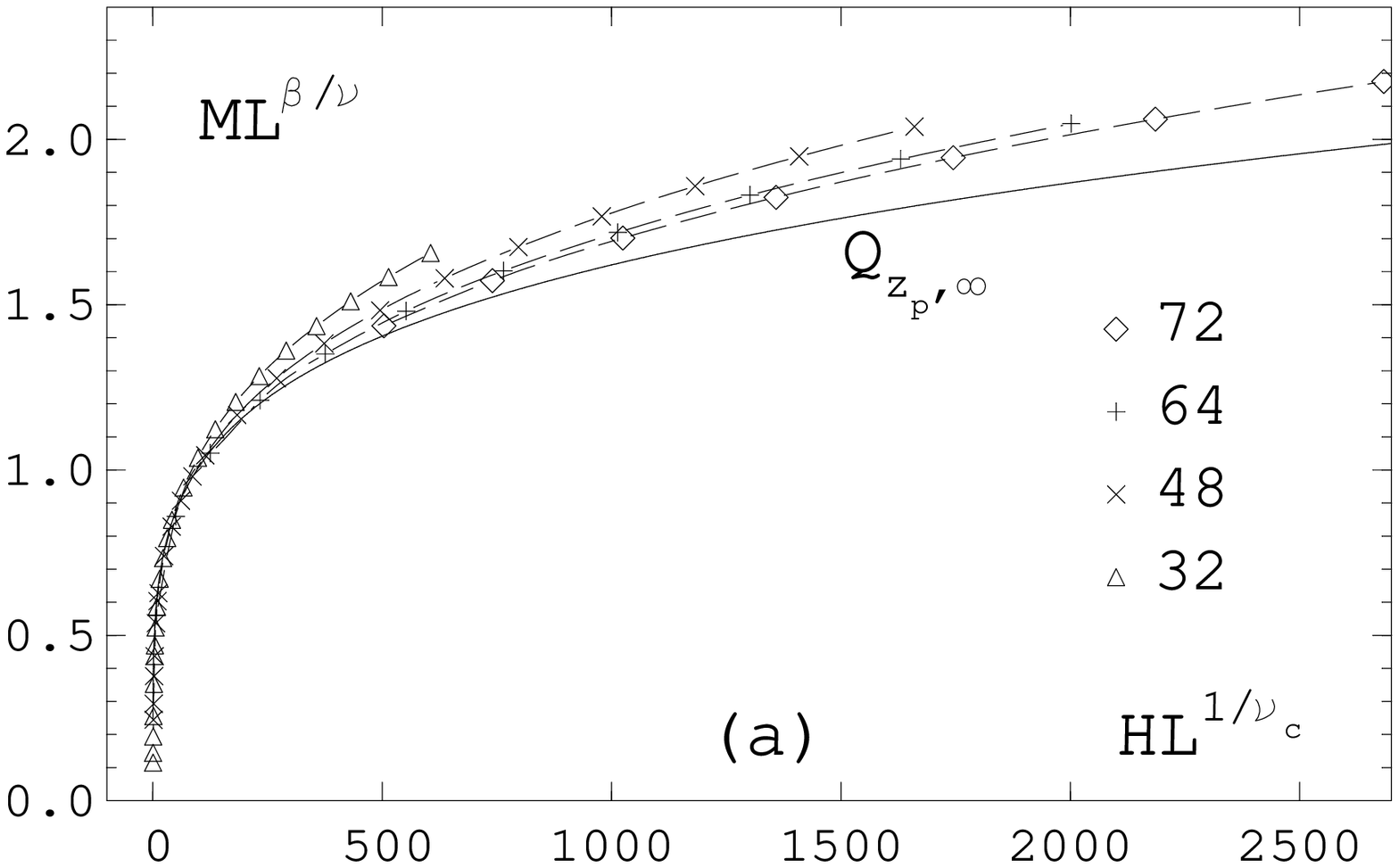, width=50mm}
   \epsfig{bbllx=40,bblly=290,bburx=370,bbury=568,
       file=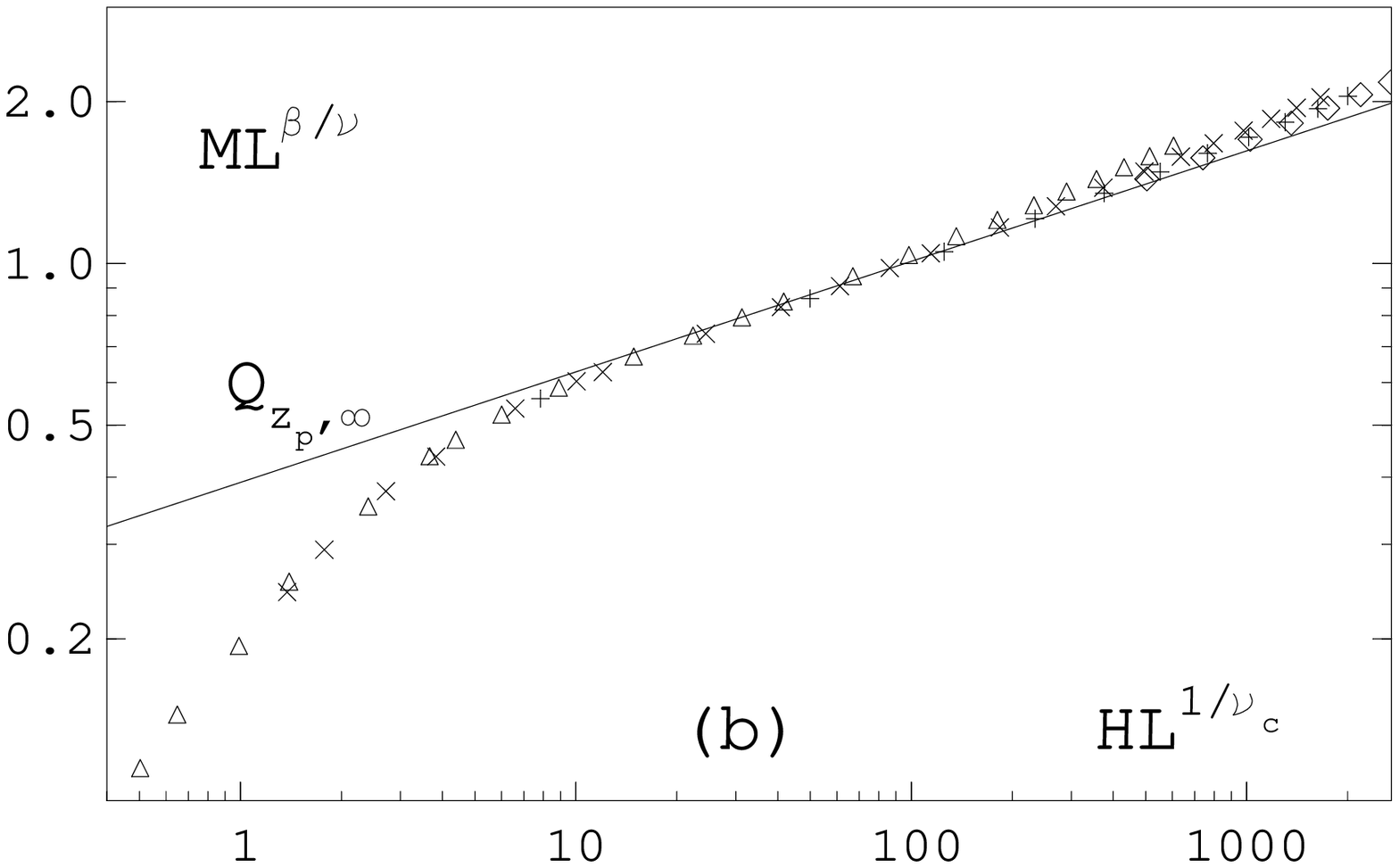, width=50mm}
\caption{(a) Finite-size scaling of $ML^{\beta /\nu }$ for $O(4)$ on the
       pseudocritical line. The solid line shows $Q_{z_p,\infty}$, the symbols
       denote different lattice sizes $L$. (b) is a double-log plot of (a).}
\label{fig:fsspseudoO4}
\end{figure}
In Fig.\ \ref{fig:fsspseudoO4}a we show the finite-size-scaling plot for
$O(4)$ on the
pseudocritical line (at $z_p=1.33$). There are strong corrections to scaling
and the universal function $Q_{z_p}$ is approached from above. If one looks at
the logarithmic plot (Fig.\ \ref{fig:fsspseudoO4}b), one finds that $Q_{z_p}$
converges to $Q_{z_p,\infty}$ from below and coincides with the asymptotic form at
about $HL^{1/\nu_c}\approx 20$.

We have also investigated the finite-size-scaling func\-tions of $O(2)$ on the
pseudocritical line and of both models on the critical line. They have a very
similar behaviour to the one shown above, but on the critical line the
corrections to scaling are negative, and in the case of $O(4)$ they are much
smaller than on the pseudocritical line. The asymptotic form is always
approached at small values of $HL^{1/\nu_c}$ from below.


\section{COMPARISON TO $N_f=2$ QCD}
\label{section:compqcd}

\begin{figure}[t!]
\begin{center}
   \epsfig{bbllx=127,bblly=310,bburx=451,bbury=587,
        file=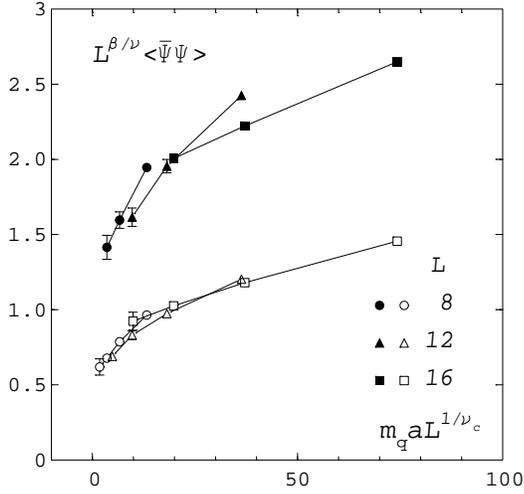,width=60mm}
\end{center}
\caption{$L^{\beta /\nu}\langle\bar{\psi}\psi\rangle$ versus $m_qaL^{1/\nu_c}$,
        with $O(4)$ exponents, from QCD lattice date with two staggered
        fermions. The open (filled) symbols are data from Ref. \cite{Kanaya}
        (\cite{Laermann}, multiplicated by 2). The lines are drawn to
        guide the eye.}
\label{fig:QCDData}
\end{figure}
\medskip
In Fig.\ \ref{fig:QCDData} we show a finite-size-scaling plot of staggered QCD
data on the pseudocritical line from the JLQCD collaboration \cite{Kanaya} and
the Bielefeld group \cite{Laermann}. The exponents are from the
$O(4)$-model. Here the quark mass $m_q$ plays the role of the symmetry-breaking
field $h$, and the chiral condensate $\langle\bar{\psi}\psi\rangle$ is the order
parameter for $m_q\rightarrow 0$. $\langle\bar{\psi}\psi\rangle$ has been taken
at the peak positions of the chiral suceptibility $\chi_m$ at fixed $m_q$ for
each lattice size.  The error from the peak position determination is not
included in the plot. As in Fig.\ \ref{fig:fsspseudoO4}a the universal
finite-size-scaling function is approached from above. At small $m_q$ the data
are however higher than expected.

\begin{figure}[t!]
\begin{center}
   \epsfig{bbllx=127,bblly=310,bburx=451,bbury=587,
        file=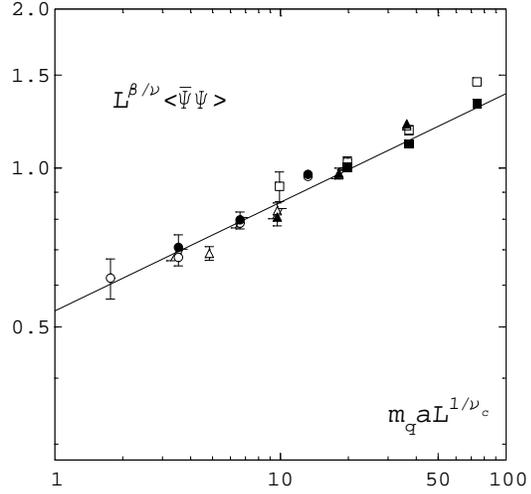,width=60mm}
\end{center}
\caption{Logarithmic plot of Fig. \ref{fig:QCDData}. The solid line is the
        asymptotic finite-size-scaling function of O(4).} 
\label{fig:Comp}
\end{figure}

In the logarithmic plot, Fig.\ \ref{fig:Comp}, we \nobreak{compare} \nolinebreak
the QCD data to a line $\bar{c} + (1/\delta )\ln (m_qaL^{1/\nu_c})$, which
represents the asymptotic finite-size-scaling function. Because of the unknown
normalization of QCD $\bar{c}$ was chosen freely. There is good agreement of the
QCD data with the expected behaviour, especially when one takes into account that
the lattice sizes are still small and corrections are probably present.

The corresponding result for $O(2)$ is very similar to that of $O(4)$ and,
because of the spread of the QCD data, one cannot really distinguish the two 
cases.

An extended version of this paper can be found in Ref.\ \cite{FSS}.

\begin{thebibliography}{99}

\bibitem{PW} K. Rajagopal and F. Wilczek, Nucl.\ Phys.\ B{\bf 399} (1993) 395.

\bibitem{CPPACS} A. Ali Khan et al. (CP-PACS Collaboration), Phys. Rev. D63
  (2000) 034502.

\bibitem{MILC} C. Bernard et al. (MILC Collaboration),\\ 
                          Phys.\ Rev.\ D{\bf 61} (2000) 054503.

\bibitem{O4} J. Engels and T. Mendes, Nucl.\ Phys.\ B{\bf 572} (2000) 289.

\bibitem{O2} J. Engels, S. Holtmann, T. Mendes and T. Schulze,
            Phys.\ Lett.\ B{\bf 492} (2000) 219.  

\bibitem{Kanaya} K. Kanaya, private communication.

\bibitem{Laermann} E. Laermann, private communication.

\bibitem{FSS} J. Engels, S. Holtmann, T. Mendes and T. Schulze, Phys.\ Lett.\
  B{\bf 514} (2001) 299.

\end{thebibliography}
\end{document}